# Art Notions in the Age of (Mis)anthropic AI

Dejan Grba

Interdisciplinary Graduate Center, University of the Arts, 11000 Belgrade, Serbia; dejan.grba@gmail.com

**Abstract:** In this paper, I take the cultural effects of generative artificial intelligence (generative AI) as a context for examining a broader perspective of AI's impact on contemporary art notions. After the introductory overview of generative AI, I summarize the distinct but often confused aspects of art notions and review the principal lines in which AI influences them: the strategic normalization of AI through art, the representation of AI art in the artworld, academia, and AI research, and the mutual permeability of art and kitsch in the digital culture. I connect these notional factors with the conceptual and ideological substrate of the computer science and AI industry, which blends the machinic agency fetishism, the equalization of computers and humans, the sociotechnical blindness, and cyberlibertarianism. The overtones of alienation, sociopathy, and misanthropy in the disparate but somehow coalescing philosophical premises, technical ideas, and political views in this substrate remain underexposed in AI studies so, in the closing discussion, I outline their manifestations in generative AI and introduce several viewpoints for a further critique of AI's cultural zeitgeist. They add a touch of skepticism to pondering how technological trends change our understanding of art and in which directions they stir its social, economic, and political roles.

**Keywords:** art notions; artificial intelligence; computational art; computer science; generative artificial intelligence

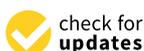

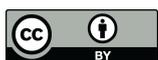

## 1. Introduction

After the emergence of text-to-image (TTI) models in 2021 (Ramesh et al. 2021) and their diversification and commercialization in the following two years, generative artificial intelligence (generative AI) entered the mainstream culture. Featuring streamlined interfaces and functionality, online TTI services, such as DALL·E, Midjourney, Stable Diffusion, Leonardo, Craiyon, and Firefly, have lowered the technical knowledge barriers for working with deep learning models that produce high-fidelity visual output and expanded the AI's creative user base beyond tech-savvy artists, artistically inclined programmers, and researchers. Their popularity prompted the incorporation of diffusion model routines into offline software and the introduction of various programming tools and techniques for multimodal media synthesis. In a volatile landscape of application frameworks, access rights, usage protocols, and filtering policies, amateurs, hobbyists, and enthusiasts as well as professional artists and studios use generative AI to produce various types of content and showcase, share, and monetize it on platforms such as Instagram and Twitter and portfolio websites such as ArtStation DeviantArt, or Behance. While some of them enter and occasionally win art competitions (Parshall 2023) and attempt to copyright their creations (Appel et al. 2023), artists whose works are used for training commercial models file lawsuits against AI companies (Schrader 2023) and seek other means to protect their work, stirring a vivid and polarized public debate. These tensions belong to the expanding web of development and deployment pathways of multimodal generative models whose breadth, economic consequences, and sociocultural implications have prompted an extensive academic investigation.[1]

However, some aspects of generative AI warrant more refined scrutiny and deeper contextualization, particularly within the context of AI's overall influence on art and creativity. AI affects art in several dimensions, most notably expressive (by introducing new topics, techniques, and forms), exploratory (through applications in art history, museology, cultural studies, and digital humanities), economic (by reconfiguring the modalities of art production and monetization), and notional (by modifying the ideas about art's character). The impact of AI on art's expressive realm, scholarly exploration, and socioeconomic status has been salient in critical AI studies (Żylińska 2020; Audry 2021; Zeilinger 2021; Cetinić and She 2022; Grba 2022a; Wasielewski 2023; McCormack et al. 2024). However, the lines and character of AI's influences on art notions are addressed less thoroughly and their discussion lacks a nuanced understanding of art's continuously changing identity brought about by the modernist avantgardes, postmodernism, and experimental art practices. They are closely related with the similarly underexplored conceptual and ideological substrate of the computer science and AI industry whose attributes the art-related AI technologies such as generative AI disseminate by facilitating the (often uncritical) proliferation of digital artefacts and fostering the integration of computational art into the mainstream culture and economy (see *Everypixel* Journal 2023).

In this paper, I take generative AI as a pretext to examine these two intertwined sociocultural facets of modern AI. Using TTIs as a leitmotif, in the following section, I make a condensed outlook on generative AI in the artmaking context. After outlining the distinct but often confused conceptual aspects of art notions in section three, I review the principal directions in which AI influences them: the normalization of AI through art, the representation of AI art in the artworld, academia, and AI research, and the mutual permeability of art and kitsch in digital culture. I frame these influences within the ideational and ideological substrates of computer science and the AI industry, whose questionable attributes propagate into culture through popular AI products such as generative AI. My main points of interest are the fetishism of machinic agency, the equalization of computers and humans, sociotechnical blindness, and cyberlibertarianism. These trends are shaped by diverse and sometimes incompatible but somehow coalescing technical concepts, philosophical premises, and political views, many of which have the overtones of alienation, sociopathy, and misanthropy. They are largely absent from or attenuated in the debates about AI's transformation of art and remain underexposed in AI studies, so in the closing sections, I summarize some of their manifestations in generative AI and introduce several viewpoints and possible directions for further critique of AI's cultural zeitgeist.

While many background problems of computer science and AI have been examined in the historical, philosophical, and sociological studies of these fields, they have not been articulated within the notional contexts of artmaking and creative expression. My goal is to do it in this paper by tracing the instrumentalization of art and creativity as the cultural normalizers of AI and the tech industry's dubious values. That is a demanding task, and it is important to recognize its limitations. The paper maps sweeping and often convoluted subjects from diverse disciplines into a compact narrative, so certain issues and their interrelations are omitted or not discussed in detail. Several indicated topics require further research and dedicated publications, such as the trends and accidents that blend individually incongruent ideas and tendencies in the computer science and AI industry, the mechanisms of their cultural impact in the areas besides artmaking, and the role of the art education sector in implicitly nurturing students' complacent handling of computational media with curricula hastily designed around hyped-up technologies. Looking at the intersections of AI and art within such a (self-)critical perspective may be conducive to cultivating an informed and responsible approach to the contemporary AI-influenced society.

## 2. Generative AI

Generative AI is a subfield of AI research that develops techniques for rendering textual, sonic, visual, and other types of digital artefacts by using models that form patterns and structures based on their training data to generate new data with similar characteristics (Totlani 2023). Multimodal generative models connect models that process the input in one medium, such as text, with models that interpret and output these processed input patterns to another medium, such as image. They benefited from the development of large models, also known as foundation models, after the introduction of OpenAI's CLIP[2] in early 2021, and often use diffusion models[3] to synthesize and improve the quality of the output. On a technical level, the realistic appearance of multimodal generative models' output surpasses the preceding media-synthesis deep learning techniques such as generative adversarial networks.

For example, TTIs use language processing multimodal models to transform text input into a latent space of image-text embeddings, and diffusion models to synthesize these embeddings into formally coherent images ([Ho et al. 2022](); [Ramesh et al. 2022]()). By combining keywords and model parameters (directives) to compose prompts, a TTI user acts as a task definer and evaluator of the resulting images, and the AI system generates visual concepts and renders their corresponding pixel arrangements. Facing the expressive challenges of TTIs' semantic constraints, users' diverse notions of visual motifs, styles, mediums, techniques, effects, and other common formal attributes have spurred a burgeoning online scene for sharing prompts, prompting techniques, and prompt-image pairs on websites such as Prompt Hero and trading them on prompt marketplaces such as PromptBase, Promptrr.io, Prompti AI, or PromptScoop.

TTIs constitute a branch of a rich ecosystem of generative AI tools and techniques, whose pronounced cultural presence and economic momentum have stirred vigorous discourse in which—uninhibited by the minuscule historical distance—the media, pundits, as well as some scholars, tend to enthuse about generative AI's disruptive power over art. For instance, [Lev Manovich]() ([2023]()) describes generative AI as a revolution comparable in magnitude to the adoption of linear perspective in Western visual arts and the invention of photography. Others believe that generative AI is a profoundly impactful medium whose "synthesis of human intuition and machine capabilities" represents a "paradigm shift" that "heralds a renaissance in artistic expression, offering glimpses into the limitless possibilities that lie ahead" ([Novaković and Guga 2024]()). They claim that generative AI transcends a mere artistic tool and makes a crucial step toward the fulfillment of the creative industries' long-standing goal to democratize artmaking into a more socially integrated and economically productive force and thus redefine the "traditional exclusivity" of professional artists' roles ([Kishor 2023]()). Wittingly or not, this effusion dances to the tune of AI science and industry, which have systematically abused the messianic rhetoric about their indispensability for social betterment in a variety of contexts, often with undesirable consequences. It faces a plethora of generative AI problems and fallouts identified in critical AI studies.

The limitations of existing TTIs make it hard to achieve the desired specificity and quality of visual output, so prompt writing amounts to an iterative trial-and-error process or guesswork. TTIs can be steered in unexpected directions with unconventional, often counterintuitive, or absurd formulations ([Bajohr 2023](), p. 67) and one of the motivating factors for using them is the excitement of combining imagination, knowledge, research, cultural browsing, and verbal gymnastics to devise prompts that may be rewarded with dopamine hits when the model's configurations of concept/pixel arrangements yield compositions that hijack our cognitive mechanism for extracting meaning from perceived (visual) patterns (see [Wilner 2021]()). TTI systems' "agencies" are thus exerted by inciting users to actively subjugate their natural language and formal concepts to the generators' artificial semantics. They emphasize human susceptibility to expressive and perceptual conditioning in interactions with technologically mediated reality ([Pasquinelli 2019](), p. 17) and present a topic for future investigation ([McCormack et al. 2024](), p. 14).

Although the synthetic surface mimicry of popular visual styles does not in itself constitute an artistic innovation, TTIs' underlying concept of turning semantically encoded images of the past into the sources of visual patterns that can be extracted and transformed beyond established hierarchies of cultural values ([Meyer 2023](), pp. 108–9) can be used for cogent artmaking. However, besides the adequate understanding of the technical process and its source material (training data), it requires "gambling" with the prediction algorithms, which usually turns out to be more interesting to TTI authors than the audience. Like other AI technologies, TTIs do not produce images of the world, but language-filtered images of, or about, other images (Ervik 2023; Meyer 2023, p. 108). Since their machine learning techniques are predictive rather than truly generative, the resulting visuals reveal various features of their models' image-text datasets as well as the human decisions behind them (Salvaggio 2023, p. 84). With TTIs, the concept of "style", broadly understood as a nameable and repeatable aesthetic mode (the formally induced "mood" or "look"), detaches from its source images, their makers, mediums, and production contexts by being encoded into the latent space data patterns pried from large online aggregators, such as Instagram, Reddit, Wikipedia, or GitHub.

In principle, the combination of linguistic with visual plasticity in TTI production frameworks is an interesting, if not unique, mode of expression. Nevertheless, its key feature—bridging the descriptive and conceptual limitations of language with generated visuals—is simultaneously its crucial weakness. That is because the medium-inherent limitations and artists' ways of overcoming them establish the resulting artworks' experiential and interpretative space in which uncertainty or vagueness play important roles but the TTIs require yielding the creative authority for transcending these limitations to an opaque algorithmic selection of the weighted averages of creative choices that have already been made. Aesthetic features must be prompted, and their renderings appear from the source material data already tagged with such features.

Furthermore, despite its impressive volume, the data corpus for TTI training does not map humanity's cultural spectrum because it is predominantly sourced from online data aggregators whose contents are dominated by the hegemonic cultures and have already been heavily moderated. In turn, most generated outputs are posted online for further scrapping, so new models' training data fuse with the existing models' output. Beyond being flawed in principle (Watson 2019, pp. 423–24) and perpetuating cultural norms already prevalent in training datasets, the regurgitative learning inflation promotes and amplifies clichés and biases, reinforces stereotypes, and widens cultural gaps (McCormack et al. 2024, pp. 3–4), possibly leading to narrow, entropic, or homogeneous new models.

The representative and epistemological powers of large generative models are significantly diminished by the systematic input/processing/output censorship of transgressive and abject contents. Complementing the regulatory policies and mechanisms that already dominate the Internet,[4] the algorithmic filtering of prompts and latent space data prevents potential "abuses" and "protects end users", but also masks the dataset's biases and model's technical flaws. This is problematic because the content management criteria are defined by AI systems owners' narrow interests and questionable competencies, for instance about what is "allowed" in an artwork. Thus, creative expression becomes neutered on multiple levels: conceptual, thematic, aesthetic, historical, and political (Riccio et al. 2024). Undesirable manifestations of human nature are inherent to culture and cultural production, so their excision from AI technologies that claim to be culturally inclusive is both (self-)deceptive and hypocritical (see Offert 2023). This contradiction also discredits the metaphor of generative AI as a "collective unconscious" (Schröter 2023, pp. 118–19) besides its highly speculative and disputed psychoanalytical underpinnings (see, for example, Mills 2018).

For all these reasons, TTIs and generative AI more generally cannot facilitate the intentions, actions, and accountabilities available to other artistic media; crucially, they do not invent, name, and further develop new concepts. After studying the character and frequency of prompts in Stable Diffusion and Midjourney, the language they cultivate, and the types of images they produce, McCormack et al. (2024) showed that TTI practices largely focus on popular topics and aesthetics, and implied that the prevalent TTI usage is a recreational activity of narrow socio-demographic groups detached from the mainstream artworld. Their findings indicate that TTIs resemble the cultural trajectories of earlier accessible AI tools for artmaking, such as the DeepDream (Grba 2023a, pp. 207–8).

## 3. Art Notions

Despite their expressive issues, TTIs spearhead the repertoire of art- and creativity-related AI technologies whose cultural spreads and economies introduce assumptions, views, and generalizations that directly or indirectly affect artistic practices, art notions, and speculations about the future of artmaking (Smith and Cook 2023, p. 1), and questions about changing art's core identity are among the most salient in generative AI debates. Acting as a subtle but powerful transmitter of concepts and ideologies from the computer science and AI industry, this multilayered sphere of influence establishes a distinct perspective for critiquing AI's cultural impact.

*3.1. Conceptual Aspects*

There are three related but distinct aspects for considering art notions: anthropological (artmaking as a human faculty and art as a cultural component), ontological (whether an artefact or event is deemed as an artwork), and disciplinary/taxonomic (types of creative practices that constitute a certain art field). While the anthropological aspect is well established and accepted, although with no agreement on art's evolutionary role (see Hickman 2016; Neubauer 2016), ontological and disciplinary/taxonomic dimensions are elusive and often get misconstrued or conflated because they are subject to interpretations that may be driven by interest or impaired by inexperience or ignorance.

3.1.1. Ontological

Artmaking can be broadly described as a set of socially bound human activities for organizing available resources (spare time, ideas, material, tools, skills) to produce and share artefacts or events that may—in variable proportions and depending on the contexts of their creation and consumption—be perceptually or mentally (aesthetically) pleasing, convey certain concepts or narratives, stimulate imagination or emotion, offer new insights, or challenge conventions, but do not need to have any pragmatic use in the conventional sense. The perceived value of an artwork emerges through an interplay between its formal and experiential attributes, the sociopolitical circumstances of its production, cultural life, and current reception, and the audience's art assessment features: intuitions, needs, interests, historical knowledge, learned references, self-reflective consciousness, understanding of the sociocultural milieu, and critical thinking (Issak and Varshney 2022).

These intricate webs of factors constantly change, so the notions of what an artwork is and the identifications of artists' social functions evade stable and consensual forms across time periods, geographies, and sociocultural strata (Luhmann 2000). Artists affect them by gradual modifications or by breaking up with professional standards, established theoretical views, and cultural norms, which in turn catalyze the notional changes about art's and artists' identities and roles. Among such transformations in the modern Western art canon—including cubism, Dada, minimalism, conceptual art, and postmodernism—Marcel Duchamp's transposition of artmaking from the reconfiguration of matter into a cognitive process of relational creativity and discovery stands out as one of the most consequential (Hopkins 2000). Duchamp eclectically blended Pyrrhon of Elis' ethics of indifference with the theories of non-Euclidean geometry and nascent nonlinear dynamic systems to establish an approach that transcends the traditional artist–object–spectator hierarchy towards a largely indeterministic meaning construction centered on the spectator's active participation (McEvilley 1988; Molderings 2010).

Duchamp's playful ideas of art as an open-ended, organic, distributed, and mutable category (Duchamp 1973) have substantially driven art's accentual shift from formal representation to a conceptual exploration that equally favors natural, artificial, physical, and imagined elements (Rosen 2022). Continuously reiterated, rediscovered, reframed, amended, and elaborated in contemporary art, they have also established the concept of distributed or fuzzy rather than stable and localized authorship. Although the modernist dictum that art defines itself (Anonymous 1969; Perloff 2010) gets relativized in a deeper historical perspective by art's social embeddedness, the self-determination model prevails in fresh and culturally unabsorbed or "rogue" creative approaches of experimental artmaking. The conceptual frameworks and vocabularies for dealing with different types of intentionality, degrees of autonomy, and other poetic attributes in the emerging arts are incessantly created, modified, and discarded.

Consequently, art appreciation has matured to appreciate that in experiencing, identifying, and evaluating an artwork, the perceiver's conceptual framework and sociocultural background are as important as the artists' creative skills, intentions, and motivations (Smith and Cook 2023, pp. 1–3). It became receptive to objects, events, or processes that do not need to be aesthetically pleasing if their combined expressive qualities facilitate meaningful communication, discovery, and learning (see, for instance, Arnason and Mansfield 2012). It requires an experientially, intellectually, and emotionally competent spectatorship attuned to the artworks' demands and, like other domains of human creativity such as science and technology, depends on the knowledge and understanding of art's historical and contemporary dynamic.

These demands are not always met in pondering art's "essences". The attempts at formulating universal art-identification algorithms in philosophy and art theory chronically fail to tackle art's open-endedness, possibly because they are mostly incited by the prevailing lack of practical artmaking experience (Penny 2017). For instance, the functional theories of art (Collingwood 1938; Langer 1953), the exhibited features theory (Ziff 1953), the institutional theories (Danto 1964; Dickie 1969; Davies 1991), or the more recent "buckpassing" theory (Lopes 2014) tend to be so dry, restricted, or vague as to convey little meaning, prove useless in real-world scenarios, become retrospectively falsified by the upcoming art practices, and ultimately get abandoned. In the realm of informal discourse, the uneven art knowledge or comprehension is often shielded behind the entitlement to judging artworks by "individual taste" usually without revealing and defending its informative qualities. In both domains, missing or ignoring the import of art's evolution and its emancipatory implications for art's flexible meaning leads to various forms of misconception about historic as well as contemporary art fields, such as AI art.

3.1.2. Taxonomic

Contemporary AI art comprises diverse practices that spring from and respond to the development of AI technologies, the expansion of AI-powered economies, and their influence on culture and society (Grba 2022a). Although the creative ideas, topics, methodologies, and presentational formats of these practices are closely related to AI research, development, and application, the types and levels of artists' involvement with AI techniques vary. The heterogeneity and mutability of poetic pathways make AI art hard to codify and categorize (Forbes 2020; Mendelowitz 2020), but the term "AI art" is commonly used because of its inclusivity for all expressive flavors contingent on AI's cultural, technoscientific, economic, sociopolitical, and historic contexts. It accommodates artworks and art practices in different areas that share important features and involve any AI technology as well as artworks that address certain AI issues indirectly.

However, AI art discourse also brims with misleading names that belong to older and broader art fields, such as "digital art", "computational art", and "generative art". The hyped-up trends in AI tech push the online, media, and pundit terminologies toward historically myopic exclusivity, so the term "AI art" currently tends to be conflated only with practices that utilize machine learning (Hencz 2023; Wall 2023) or generative AI technologies (McLean 2024), or just with AI-created visuals (Wikipedia 2024a), while Adobe (2024) "teaches" us even more specifically that AI art is the imagery produced with their TTI model Firefly. On the TTI portfolio websites, the general descriptor "art" has been casually associated with predominantly display-presented digital images which represent a miniscule fraction of artistic techniques, forms, and media.

This taxonomic drift is well illustrated by the heavy misuse of the term "generative art". Generative methodologies are media-independent and include heterogeneous creative approaches for consciously and intentionally interfacing the predefined systems with different factors of unpredictability in preparing, producing, or presenting the artwork (Galanter 2016). Generative artwork's poetic value often depends less on its formal aesthetics than on its capacity to convey (explicitly or intuitively) the artist's cognitive process(es) in devising the logic and mechanism for interrelating controllable and uncontrollable elements with a chosen and equally important conceptual framework (Memelink and van der Heide 2023). However, even before it became mainly associated with generative AI, the term "generative art" had been conflated with earlier computational practices that involved randomness, complexity, and machine learning (Benney and Kistler 2023). Foregrounding the creative uses of currently vogue technologies, such AI art vocabularies reduce the space for the appreciation of a complex art field with strong scholarship and deep historical foundations to marketing labels and promotes its uncritical appreciation.

*3.2. The Cultural Normalization of AI*

The history of corporate attempts at culturalizing computer technologies through art and creativity dates back to the late 1950s (Slater 2023). TTIs and other generative AI products continue the series of stabs with AI technologies, such as DeepDream and style transfer apps/services (since 2014), ArtBreeder (launched as Ganbreeder in 2018), and Runway (since 2018), accompanied by AI industry's programs for art production/showcasing, such as Google's Artists and Machine Intelligence (since 2016) (Agüera y Arcas 2016) and Google Arts and Culture (since 2011) (Wikipedia 2024b). We should note that the introduction of AI-powered tools for artmaking was heralded and aided by open-source and academic projects, such as Rebecca Fiebrink's Wekinator (since 2009), a machine learning program that translates body gestures into interactive systems without manual coding (Fiebrink 2009), or a software named The Painting Fool, developed in 2012 by Simon Colton and Computational Creativity Group at Imperial College in London (Colton 2012).

However, the popularization of AI techniques for artistic purposes, such as generative adversarial networks, and the marketing of consumer-grade AI tools designed specifically for artmaking, such as TTIs, have never been just casual byproducts of AI's evolution. Releasing attractive user-friendly devices for creative expression and supporting professional AI art production (Zeilinger 2021, p. 13) benefit the AI industry's marketing, development, and public relations as widely adopted products become "indispensable", provide beta testing feedback and learning data from a large user base, and help associate AI with unique human faculties such as artmaking. Leveraging the tradability of digital artefacts facilitated by the art market's integration with crypto finance in the late-2010s (Quaranta 2022), generative AI has surpassed the AI industry's earlier attempts at cultural normalization (Sanchez 2023; Statistia 2023). It has expanded the public acceptance of AI art and its superset computational art to a degree that earlier generations of artists in these fields could not have dreamt of reaching for decades.

How exactly these socioeconomic developments have benefited computational art after its long and often troubled dwelling on cultural margins (Taylor 2014) is an unsettling question. With the so-called post-digital and mainstream AI art, artists collapsed malleable digital data into non-digital or non-interactive media to render conventionally tradeable works (Paul 2015; Grba 2023b, p. 68). With the crypto art, they embraced the imposition of false rareness and non-fungibility on intrinsically shareable and mutable digital artefacts (Grba 2023b). With the generative AI, the most prevalent and lucrative practices privilege figurative plastic motifs in popular genres of "surreal" or fantasy art, game design, comics, anime, or illustration, and a fixation on surface aesthetics and stylistic norms at the expense of other poetic factors (McCormack et al. 2024). For each gain of cultural acceptance and economic success, computational artists have yielded some of their field's intrinsic features to the power of capital and accepted the price of encouraging the production and consumption of aesthetically pleasing, desirable, and "collectible" artworks, which may resurge the simplistic notions of art.

*3.3. Authorship and Agency*

Critical AI studies identify TTI-generated images as artworks that lack an acceptable author figure, or discard the prompt-makers as artists and assign (conjunctive) authorial identity to the echelons of scientists and programmers who designed TTI tools, or consider TTI tools as artworks but not the images they produce (see Wilde et al. 2023, passim). In many respects, these views reshuffle the older contestation points about the computational art authorship that affected AI art as soon as it gained cultural prominence in the late 2010s.

The malleability of notions such as agency, authorship, and originality was central to modernism and computational artists have addressed it since the 1960s, but the specter of misidentified creative autonomy has always been haunting their efforts because sophisticated computing technologies can easily trick us into conflating human creative decision making with its highly formalized emulations. Anthropomorphic legacy has been endemic in AI art from the pioneering work of Harold Cohen who flirted with the "creative serendipity" of his painting/drawing robot AARON (1971–2016) (Grba 2023a), through mystifications about "the blurring line between artist and machine" (Elgammal 2018; Miller 2019a), to the emotionally charged claims about the agency of machine learning programs (Audry 2021, p. 85). The art market has capitalized on this rhetoric as the allusions of "exotic" artworks produced by expressively motivated AI systems rather than by humans who deal with AI technologies leverage the momentum of the ongoing AI hype (Epstein et al. 2020; Browne 2022). It boosts artists' "immunity" to the criticism and debunking of such notions (see Browne and Swift 2019; Browne 2022), and they frequently discuss their AI devices as "creative collaborators", "partners", or "companions" (Audry 2021, pp. 27–28, 241–43). The audience's romanticized, anthropomorphically skewed perception and virtue signaling about AI artworks reinforce the myths of machinic agency, which in turn encourages artists and the art market to exploit them further (Ruff 2022; Grba 2022a, pp. 3–5; 2024, pp. 3–6).

Some critical AI studies join this feedback loop by metaphorizing the TTI image generation processes as forms of the artists' externalized visual cognition (Feyersinger et al. 2023) and generative models as co-creative agents (Scorzin 2023, p. 189). The related claims that the death-of-the-author effect (Barthes 1967) and the accumulation of human creative contributions in complex artmaking tools justify (and make meaningful) the assignment of authorship to inanimate entities such as generative AI systems disregard that, within such line of reasoning, both the cultural inheritance and technological accumulation of creativity equally "dispossess" AI systems and human artists of authorship. A more coherent take on generative AI regards it as a sophisticated remediation apparatus related to earlier remix techniques because generative models depend on a predictive amalgamation of sampled artefacts whose formal features are insinuated by the output imagery (Smith and Cook 2023, p. 2; Bolter 2023).

The metaphors of AI systems' artistic autonomy in art discourse simultaneously rely upon and dismiss the heteronomy of digital technologies, the centrality of sociocultural relations in their uses for artmaking (Verdicchio 2023), and the vitality of human wit and ingenuity. In AI research and industry, this contradiction manifests in the coexistence of representing AI (built by human creative efforts) as an augmentation and redemption of feeble or exhausted human creativity and treating advanced AI systems as autonomous creators. It is exemplified by the persistence of the Turing Test variations in computational creativity (CC) studies[5] whereby subjects identify or assess human-made against AI-produced artefacts under controlled conditions (see Moffat and Kelly 2006; Ragot et al. 2020; Daniele et al. 2021) despite the disputed conceptual clarity of the Turing Test and the dismissal of its relevance for detecting intelligence (discussed in more detail in Section 4.2.1).

Some CC experiments successfully show how AI algorithms can produce novelty but tend to falsely equate novelty generation with artmaking, struggle to account for the intentionality and sociocultural embeddedness of making and appreciating art, disregard the inherent artificiality of artworks, and miss other subtleties of their exploratory subjects.

They often ask inappropriate or misleading questions ("Who is the [real] artist?" or "Which artwork is better?"), but ignore or exclude participants' demographics and art evaluation proficiencies, and underestimate the power of cultural cognitive maturation that coevolves with pervasive technical systems such as AI.[6] The baffling disparity of CC studies' findings demonstrates the researchers' conceptual or methodological inability to handle the central points of art ontology. It is one of the indicators that the overall artistic literacy in AI science and industry is a product of self-assured art dilettantism informed by extra-disciplinary scholarship where art-savvy technocrats and affiliated tech-savvy artists play important roles, which may be suitable for some art-related research areas but does not map the range of practical and theoretical insights across art fields.[7] This mismatch between artistic aspirations, competencies, and outcomes has been historic in computer science/tech.

### 3.4. Tricky Distinctions

Various ideas of turning computers into handy multimedia machines that aid human creativity have been floating in computer science since Vannevar Bush's Memex in 1945 (Bush 2003). Alan Kay and Adele Goldberg articulated them with the influential concept of "active metamedium" in a discussion of their 1977 Dynabook project that paved the way for the modern laptop (Quaranta 2023, pp. 212–13). Since then, computer science and tech industry have steadily brought about novel methods for digital content creation[8] and reinvigorated older ones (such as bricolage and remix), with convergent cultural consequences. From user-friendly apps to programming languages with steep learning curves, computational artmaking tools translate certain (not all) features of previously established artistic media or techniques and often open new possibilities. They usually allow programmability or extensibility, but inevitably miss some poetically decisive factors of technological decision making inherent to their analog source art practices and feature different motivations and affects. The expressive routes, conceptual values, and aesthetics of digital media are further influenced by the tech designers' (rarely neutral) practical and aesthetic ideas about their sense and purposes, the trade-offs and compromises in the research and development processes, the production standards/frameworks, and the legislative concerns, economic interests, and political views that shape the end-products' interface metaphors and operational protocols (Winner 1997, p. 14; Fuller 2008).

When recognized, these impositions on digital media may stimulate sophisticated creative thinking. However, most early adopters are overwhelmed by meeting the new tools' cognitive demands and more preoccupied with exploring their exciting capabilities than studying their art-historical and techno-cultural backgrounds, which leads toward formalism and technocentrism (see Grba 2023a, pp. 217–18). Once widely adopted, digital tools' expressive conditioning remains mostly hidden behind typical usage scenarios and conventional practices that predispose trivial or uncritical approaches (Quaranta 2023, pp. 213–14, 216).

3.4.1. Porous Perimeters

Before the cultural prevalence of digital media, the expansion of post-World War II mass-fabrication and reproduction technologies, consumerism, and the growing volume of modernist art had led to a formal and semantic saturation, which was one of the key incitements of postmodern art. It emerged between the late 1960s and late 1980s as a heterogeneous corpus of tendencies in literature, visual arts, music, cinema, and architecture, which contested or contradicted some features and principles of modernist art and explored the sense of expressive/authorial crises in the inflated but increasingly homogeneous global culture. Postmodernist art is broadly characterized by collage, bricolage, appropriation and historicism, stylistic and thematic recycling and remix, the use of text in visual media, simplification, and intentional indiscrimination between "high" arts and popular culture (Wallis 1992; Butler 2003). Most of its features are ostensibly shared by generative AI art, but an instructive way for comparing these two fields' contextual and expressive parallels is by comparing Art & Language's paintings in the *Portrait of V.I. Lenin in the Style of Jackson*

*Pollock* series (1980) (Artsy 2024) with the outputs of popular TTIs prompted with a "portrait of V.I. Lenin in the style of Jackson Pollock".

Postmodernism catalyzed the artworld's diffusion—which had been steadily growing throughout the 20th century—into a hybrid multicultural conglomerate of ideas and approaches, with many centers and as many peripheries where artists often evade codified roles and engage in revisiting and mixing genres and styles. Artworld's later absorption of accessible digital technologies has further dissolved the previously set distinctions between professional and nonprofessional content producers and, after the introduction of blockchain technologies in the late 2000s, these changes have encouraged hobbyists, prosumers, amateurs, and weekend artists to compete in an attention economy that often turns them into commodities (Quaranta 2023, pp. 207–8, 211). At the same time, the art market has started sharing its selection criteria for computational art with crypto investors acting as art collectors and with monetization algorithms on NFT trading websites (see Quaranta 2022, pp. 221–27; Grba 2023b, p. 68). This confluence has fomented a straightforward, mostly automated proliferation and monetization of eye-catching digital artefacts (crypto art) whose styles converged toward derivative platform-powered aesthetics akin to zombie formalism (Wiley 2018a, 2018b; Hegenbart 2019). Walter Robinson (2014) used the term "zombie formalism" to criticize the commercially driven resurgence of abstract paintings in the mid-2010s, produced with reductive, essentialist techniques to look simultaneously elegant and suggestive of the elemental materiality of painting. Zombie formalism was an example of "artistically literate" kitsch (competently produced trivial artworks) which is historically integral to art and exists in a grey zone between experimental practices and plain kitsch.

Zombie formalism was abstract, crypto art comprises both abstract and figurative visuals, while the TTI imagery is largely figurative and dominated by an even more pronounced kitschy look. The perpetuation of cultural norms, biases, hegemonies, stereotypes, and injustices makes the TTI production a conceptual antipode of art brut—art created by individuals operating beyond the official cultural boundaries (obscure amateurs, psychiatric patients, prisoners, etc.) and distinguished by uninhibited freshness, noncompliance to expressive canons, and disregard of conventions imposed by professional training.[9] Like zombie formalism and crypto art, TTI production serves as a reminder that art's notional open-endedness equally applies to kitsch as "art's shammy doppelgänger" in an affair that has become more intense, promiscuous, and simultaneous.[10] As Robinson (2014) put it, "The notion that there is a genuine, pure, sincere, and deep art that can be set in opposition to a compromised, mercenary, dishonest, and shallow one is a romantic piffle".

3.4.2. Critical Approaches

An excellent vantage point for contemplating the intersections of art, kitsch, and (generative) AI opens with Vitaly Komar and Alex Melamid's project *The People's Choice* (1994–1997). In a shrewd take on using statistics for artwork "optimization", the artists hired professional polling companies in fourteen countries[11] to survey 1001 local adult citizens about their favorite themes, shapes, and colors in painting and used the results to make a Most Wanted and a Least Wanted painting for each country (Komar and Melamid 1997). The kitschy pungency of the results—realist medium-sized autumnal landscapes with animals and people in the Most Wanted series and small, hard-edge geometric abstractions in the Least Wanted—makes *The People's Choice* a satirical but also, as the artists legitimately insist, a sincere (averaged) rendering of popular tastes (Dia Center for the Arts n.d.).

By ridiculing the idea of art modeled on polling techniques regularly used in marketing and politics, and by discrediting the libertarian belief that the "wisdom of the markets" can make art production more "scientific", "democratic", or "accessible", *The People's Choice* also warns about the pitfalls of prescribed directives for art's identities and social roles. As James Dickinson (2001, pp. 150–51) noted, despite its jokingly stated aim to escape the art/kitsch discrimination by listening to the voice of democratic states' "silent majorities", *The People's Choice* ultimately (and knowingly) reproduces it. That is because the initiatives to "democratize" art tend to forsake key (and hardest) requirements of professional artmaking—cultivated sensibility, dedication, and meaningful motivation—in favor of expanding the accessibility of techniques for producing, presenting, and trading artefacts. Thus, after speculating about paint-by-number kits and computer programs as alternative routes to democratize art, Dickinson discarded them as populist. He could have been writing about the TTIs twenty years in the future.

Contemporary artists' critique of generative AI at this point largely articulates in reactions to the misappropriation of their creative efforts, particularly in lawsuits against companies such as OpenAI, Meta, Google, and Stability AI for scraping their copyrighted material and personal information to train generative models without consent or compensation. Tech companies offer them half-hearted or semi-effective solutions, such as the option to "opt out" by filing a request that their data are not to be used for AI model training. However, the option only applies to excluding artworks from the upcoming, not the existing TTI models, and even after a requester completes a tedious process of removing images from training datasets, there are still no real guarantees that they will not be used for training (Kapoor and Narayanan 2023). Here, the corporate AI's hegemonic position manifests in the cynical treatment of art stakeholders' rights, which primarily functions as a whitewashing strategy and a preemption of future litigation. Thus, artists increasingly rely on an emerging research area called machine unlearning, which develops "training data poisoning", "image cloaking", and "style masking" techniques that can protect chosen data patterns from being exploited by large learning models. They use software such as Nightshade (Shan et al. 2023a), My Glaze (Shan et al. 2023b), or Art My Choice (Rhodes et al. 2023) to add human-invisible pixel patterns in their images, which cause generative models to break in unpredictable ways and could damage their future iterations (Salman et al. 2023). The contaminated training data are currently hard to remove, but it is reasonable to expect the emergence of robust data poisoning defenses (Heikkilä 2023).

Artists also repeatedly demonstrate that any technology can be leveraged for sophisticated critical expression beyond pragmatic defensiveness, and one possible direction with generative AI could be working against the currently prevalent concealment of the digital/synthetic nature of TTIs' output (McCormack et al. 2023, 2024). TTIs can also be used more radically to obstruct content legibility, determinacy, and clear interpretation, which undermines generative AI's posing as a communication-based creative interface that simulates meaning on its side (see Bender et al. 2021). For instance, Jake Elwes' *A.I. Interprets A.I.: Interpreting 'Against Interpretation' (Sontag 1966)* is a three-channel video installation that exploits the mutual I/O feedback between two AI programs (Elwes 2023). An image-generating diffusion model (Disco Diffusion) is prompted with sentences from Susan Sontag's seminal essay Against Interpretation (displayed in the first channel) to produce images (second channel) that are then interpreted back into language by the GPT2 and CLIP image labeling system (third channel). With a bizarre authoritativeness of the resulting, mostly meaningless or misinformative reinterpretations, this work emphasizes (generative) AI's notional reduction of both visual and narrative arts, analogous to Robert Morris' *Self-Portrait (EEG)* (1963) that critiqued the neuroscientific reduction of human mind/consciousness to the measurable brain functions (Krauss 1994) and Marc Quinn's works such as the Self-Conscious (2000) that critiqued the reduction of human psychophysiology to the genetic code (Quinn 2000).

However, even if critical of kitsch, art can foment it through the hasty content proliferation or the introduction of ideas and techniques susceptible to commodification and recuperation (Grba 2022b, pp. 68–69). Acknowledging these pitfalls together with historically unstable, porous, and elusive but still identifiable art/kitsch demarcation lines, Domenico Quaranta (2023, p. 222) argues that the artworld, despite all its imperfections, should keep its "gatekeeping" role as an interactive consensus mechanism for determining art's cultural values. It seems, however, that artworld's heteronomy (Stallabrass 2006), unresolved contradictions, and inherent social injustices (Grba 2023a, pp. 217–18) have relegated the responsibility for conceiving art notions to the knowledge and critical thinking of art-interested individuals. The erosion of the artworld's and art market's arbitrage entitlement is exemplified by their repeatedly compromised relationship with computational art before the 2010s, while the coincidence of computational art's subsequent "rehabilitation" with the NFT- and generative AI-driven eruptions of kitsch indicates these institutions' economic complacency rather than informed intuition about art's poetic and cultural values (see Grba 2023b).

## 4. Undercurrents

Parallel to inciting apparent changes in creative production trends, generative AI impacts art as a transmitter of ideas, interests, and inclinations from mainstream AI. A collection of diverse and sometimes incongruous but somehow coalescing technical concepts, philosophical premises, political views, and ideological tendencies in AI science, technology, and industry wields a strong if seemingly indirect influence on contemporary mindsets. Many forces in this flux have the overtones of alienation, sociopathy, and misanthropy, which, due to active avoidance or genuine incognizance, largely escape the debates about AI's transformative effects on artmaking and society. Nevertheless, they infuse culture, translate into art practices, and shape the professional and popular notions of art, creativity, and AI. Here are some of them.

### 4.1. The Machinic Agency Fetishism

Human intelligence has been the most prevalent source of inspiration for AI design and many AI techniques either deliberately or coincidentally mirror certain aspects of human cognition to varying degrees. However, it is often hard to evaluate the difference between the effectiveness of human intelligence and the efficiency of task-specific artificial processes related to the concepts of human intelligence because of anthropomorphism—an innate psychological tendency to assign human cognitive traits, emotions, intentions, or behavioral features to nonhuman entities (Hutson 2012). Human–machine intelligence analogies and metaphors are both tempting and elusive, so anthropomorphism permeates the foundational concepts, terminology, and notions of intelligence in AI science and industry as well as popular culture (Salles et al. 2020). Throughout the history of computer science, the epistemological and metaphysical confusions caused by conflating human intelligence and machine performance have rendered anthropomorphism and AI inseparable, and some authors suggest that it is more feasible for AI research to manage anthropomorphism than strive to purge it (Proudfoot 2011).

Of course, there is no reason for conflating a nonliving system with a biological entity just because both can perform certain functions that are computationally interpretable. However, the media and some AI scientists frequently associate the performance of state-of-the-art AI systems with human cognitive traits such as intuitive physics, intuitive biology, intuitive psychology, causal models, active social learning, conceptualization, subconscious abstraction, generalization, analogy making, and common-sense reasoning—the very capabilities these systems lack the most (Mitchell 2019). As Browne and Swift (2019, p. 3) pointed out, in the language of AI, assertions that a machine "learned", "discovered", "outsmarted", etc., presuppose agency and often imply consciousness, but even placing a machine as the subject of a sentence is dubious and deserves examination. AI design and deployment is part of a socially constructed context in which humans deliberately outsource certain tasks to machines, but anthropomorphism implicitly grants them a degree of agency that overstates their true abilities, which can have profound material and ethical consequences in high-risk and sensitive application domains. Crucially, the transfer of operational authority to algorithms does not absolve humans of responsibility (Watson 2019, pp. 417–40).

### 4.2. The Objectivization of Humans

In addition to anthropomorphism, many problems of AI development and application arise from awkward understandings of computers vis-à-vis humans and paradoxical tendencies toward their mutual equalization reaching back to the foundations of modern computer science. They make AI an important constituent of the techno-cultural and social dynamics which turns a what into a who and vice versa through fallacious mental processes, flawed individual practices, and malicious institutional policies.

4.2.1. Computers as Humans

One of the unfortunate consequences of Alan Turing's legacy is the intentional or accidental provision of a "scientific basis" for the equalization of human beings and computers. In his paper On Computable Numbers, With an Application to the Entscheidungsproblem (Turing 1936), Turing first described an "automatic machine", which was later named Turing machine and became one of the key concepts in computer science. The paper was published in 1936, before the advent of automatic computing, when many people in business, government, and research establishments professionally carried out numerical calculations. These human calculators were called "computers" and Turing reemphasized in various forms that the terms "computation" and "computable" in his paper refer to an idealized description of their work (Copeland 2020). Thus, Turing's analogy between a highly structured set of operations performed by human beings and idealized computing machines makes sense only within the specific historical and utilitarian contexts of his writing.

However, he ostensibly went from connecting the isolated features of human and machine computation toward conflating human beings with computing machines. In a 1950 paper Computing Machinery and Intelligence (Turing 1950), Turing proposed the Imitation Game as a method for testing a computational machine's ability to exhibit intelligent behavior equivalent to, or indistinguishable from, a human. It became known as the Turing Test, and this name has often been associated with other types of tests for the presence of intelligence in artificial systems. However, Turing centered the proposal around an unclear concept of intelligence and left many other parts of the discussion open to interpretation, causing a long-lasting controversy (Hayes and Ford 1995; Moor 2001; Levesque 2017; Oppy and Dowe 2020).

The methodology of assessing human intelligence through a single, highly formalized channel of linguistic communication (written text) is too narrow as thinking is frequently nonverbal and combines verbal and nonverbal mental processes with numerous other factors (Tulio 2021). More general objections posit that Turing devised the Imitation Game aiming to legitimize the so-called "null hypothesis" of no behavioral difference between certain machines and humans, which was arrogant because it assumed understanding human cognition without first obtaining a firm grasp of its basic principles (Searle 1980; Block 1981). Turing's affinity for the "null hypothesis" also provides grounds for an argument that aloofness, narcissism, and psychological issues evident throughout his life "conspired" to elicit a misanthropic bitterness, which motivated the infantile computer–human analogy.[12]

Since Turing, the range and persistence of grotesque notions in AI research indicate both conceptual and mental issues so, as Browne and Swift (2019, p. 2) noted, it is important to acknowledge their connotations and consequences:

> The separation between "reasonable" and "unreasonable" ideas [in AI science], which we might call superstition is less clear than one might expect. In Computing Machinery and Intelligence, Alan Turing considers the use of a "telepathy-proof room" to protect the integrity of his Imitation Game from players exhibiting extrasensory perception. This may cause us to cringe in hindsight—it's uncomfortable to imagine heroes of science believing such unlikely things. But good science demands open-mindedness and the courage to challenge accepted truths. AI researchers are in a difficult position, expected to dismiss "silly" ideas like telepathy and yet take seriously the idea that bits of metal and silicon might become intelligent if you program them the right way.

Despite the understandable challenges of cutting-edge thinking, the AI community's leniency toward its members' quirks[13] is regressive and irresponsible. The notion of a personified computer awards the nonliving entity the role of the Other, places it into our circle of empathy (Singer 2011), and assigns it elevated rights while we have long been surrounded with living "candidates" for expanding our empathy and improving our ethics but still do not treat them consistently and justly: other human beings, animals, and plants. The consequent logic of personified AI implies that we need to devise value systems—urgently needed but inadequately applied to many existing beings—on a purely speculative model of sentient machines.

How do such ambiguities translate to the reality in which the rapid industrialization and widespread application of AI technologies bring about the concentration of wealth and political power that leads to a society contingent on corporate AI interests?

4.2.2. Sociotechnical Blindness

Systems that involve frequent information exchange and processing can, for some practical purposes, be envisioned and treated as data structures. Thus, data collection and quantization, behavioral tracking, predictive modeling, and decision-making manipulation have long been essential strategies for large-scale information-dependent systems such as governments, industry, marketing, finance, insurance, media, and advertising. By coupling massive digital datafication with sophisticated statistical algorithms, modern AI increases the extent, intricacy, and efficacy of these social engineering strategies, whose undesirable effects arise from the disparities between business priorities (maximizing wealth and competitive power), the impact of AI products on various demographic groups, and broader societal interests (O'Neil 2016; Zuboff 2019).[14]

Continuing the tradition of using human beings as hidden micro-components in large computational architectures since the late 19th century, the AI industry devises algorithmic methods and frameworks for simultaneous large-scale data collection and processing, productivity maximization, and workforce concealment. It often applies legally dubious labor policies and unethical human resources management practices that evade control and regulation although they have been thoroughly documented (Lorusso 2020; Zukalova 2020). Since data harvesting operations pioneered by Google in the 2000s and online scraping practices starting in the mid-2010s, AI development has also ridden, and abused, a razor-thin line between research and commerce. For instance, in a process called "data laundering" or "data washing", AI companies exploit academic and non-profit projects to gather data for training models that soon turn commercial (Jiang et al. 2023). In aggregate, these trends contrive an illusion that human-created and human-dependent AI systems have high levels of material abstraction and functional autonomy. It has been identified as "sociotechnical blindness" (Johnson and Verdicchio 2017), "fauxtomation" (Taylor 2018), "ghost work" (Gray and Suri 2019), "human in the loop" complex (Paulsen 2020), and "banal deception" (Natale 2021).

The human labor demands for generative AI development have somewhat changed from their role in trailblazing the preceding AI techniques and pipelines (Williams et al. 2022) but remain excessive and harmful. Building large/foundation models requires massive human-judgment-based work on data after pre-training stages, euphemistically called "post-training alignment" in the AI literature. It includes identifying, comparing, selecting, classifying, and rating various types of data (scraped, AI-generated, or manually entered), writing examples for the model's preferred behavior (output to certain types of input/prompts, questions followed by correct answers, descriptions of computer programs followed by functional code),[15] and assembling large online media agglomerates and datasets such as LAION-5B. Performed by online workers, outsourced workers in the global South, workers in start-up platforms such as ScaleAI, industry-affiliated academic institutions, the base workforce stack of the AI industry, and by the end users, many of these tasks are repetitive and meaningless, and labor conditions are precarious and surveilled (Dzieza 2023; Solaiman et al. 2023; Beetham 2023; GlobalData 2023).

*4.3. Ideologies*

Exploitative datafication is certainly not the most impressive achievement of AI research but is emblematic of corporate AI's social politics. Since the mid-1960s, the worldviews in computer science communities and IT industries, particularly in the US and other anglophone countries, have been shaped by a bizarre ideological conglomerate of doctrines, such as utopianism, counterculture, individualism, libertarianism, and neoliberal economics (Turner 2008; Gere 2008; Rushkoff 2022). This assemblage, also called the Californian ideology (Barbrook and Cameron 2008) and cyberlibertarianism (Winner 1997), comprises ideas fueled by the zeal for technologically mediated lifestyles and future visions steeped in libertarian notions of freedom, social life, and economics. It promotes technological determinism (Mickens 2018) and techno-solutionism (Morozov 2013), radical individualism, deregulated market economy, trust in the power of business, and disdain for the role of government (Payne 2013; Armistead 2016). These values fully make sense only within the context of the right-wing political milieu and are often spiked with radical pseudo-philosophical rhetoric, such as Objectivism (McGinnis 2012; Robephiles 2022; RationalWiki 2023), that provides the "intellectual authority" for greedy technocracy.

The cyberlibertarian tendency to conflate social and political with technical problems can be summarized in the three assumptions of technological manifest destiny: (1) technology is apolitical so it will automatically lead to good outcomes for everyone; (2) new technologies should be deployed as quickly as possible, even with incomplete knowledge about their functioning and societal impacts; and (3) the past is generally uninteresting and history has nothing to teach us (Mickens 2018). After the introduction of blockchain technologies, the cyberlibertarian techno-solutionist politics has been radicalized by the burgeoning start-up community of predominantly white male crypto entrepreneurs obsessed with quick success and tending toward sexism, racism, misogyny, homophobia, and transphobia (UNESCO 2020). Most of the early developers and influential adopters of cryptocurrencies belong to intersecting movements that include cypherpunks, crypto-anarchists, transhumanists, Singularitarians, Extropians, self-described hackers, open-source software developers, and tech-savvy entrepreneurs. Leading crypto investors, such as Elon Musk, Peter Thiel, Eric Raymond, Jimmy Wales, Eric Schmidt, Saifedean Ammous, and Travis Kalanick, openly adhere to libertarianism and Objectivism, and promote economic views that range from the Austrian and Chicago schools of economics to the Federal Reserve conspiracy theories (Golumbia 2016, pp. 10, 30, 60). Therefore, despite the nominal commitment to widely acceptable social values, many crypto-economic and IT ventures epitomize pivotal right-wing politics. Cyberlibertarianism thrives behind the AI industry's facade of objectivity, rationality, progress, and political correctness, whereas its reality is dominated by aggressive competitiveness within an adversarial business culture that reflects the most unpardonable tenet of capitalism: prioritizing profit over people (Wiener 2020). AI industry values "uniquely human" skills such as attention, care, critical judgment, taste, imagination, improvisation, spontaneity, sincerity, empathy, intimacy, and humor not because they evidence individuality or authenticity but primarily because they cannot be automated for generating surplus value (Horning 2015; Gosse 2020).

While some authors deem such logic morally untenable and destructive (Eubanks 2018) and others remain undecided (Epstein et al. 2023, pp. 8–11) or claim the opposite (Kalish and Wolf 2023), it is worth remembering that, insofar as we take advantage of AI's sociotechnical regime, we share a degree of responsibility for its existence and consequences. This entanglement is evident in the ethical inconsistencies of some leading critics of technocapitalism who selectively enjoy certain layers of its gravy train by patronizing convenient services of companies that epitomize the most acute points of their critique, which may be interpreted as unprincipled or hypocritical. For instance, authors such as Shoshana Zuboff, David Golumbia, and Paris Marx choose publishers who sell their books on Amazon.com rather than less lucrative alternatives, such as the Institute of Network Cultures (INC), which allows readers to either purchase INC books on their website or download them for free.[16]

The suppression of AI technologies' social rootage, the concealment of human roles behind their performative power, the filtering of human benefits from using them, and the misrepresentation of human interests in the social conflicts they foment lurk behind modern AI's economic triumphs (Golumbia 2009, 2015). As Martin Zeilinger (2021, pp. 12–13) observed, true to their origins in the military–industrial complex, the implementations of mainstream AI are aligned with capitalist ideological frameworks and socioeconomic regimes that rely on automation, high-speed calculation, data-intensive analytics, and computationally afforded prediction. In such context, generative AI technologies can be seen as a forefront of the reiterative entrepreneurial process toward emancipating capital from humanity (Dyer-Witheford et al. 2019, p. 7), in which human work and data provision are exploited to build systems that automate certain tasks and reconfigure humans' roles for the next iteration. It decreases workers' longer-term well-being as their abilities and skills increasingly become obsolete and redundant, forcing them to maintain relevance by improvising or retraining for new competencies (Barley 2020).

## 5. Discussion

While these sinister undercurrents have been explored in the historical, philosophical, and sociological studies of computer science and AI, they require wider attention in artistic communities because the AI industry instrumentalizes art and creative expression for the cultural normalization of problematic values in its background. These values are largely obscured or attenuated in the debates about AI's cultural impact and remain underexposed in AI scholarship.

### 5.1. Generative AI as a Cultural Conduit

Most art technologies are either gradually developed by artists or co-developed between artists and scientists/engineers or introduced for non-artistic purposes and artists discover their expressive potentials through experimentation. In principle, these types of circumstances allow artists to nurture a viable critical distance toward the moral, socioeconomic, and political issues of their technological environments (Winner 1980). Nevertheless, the reality is dominated by the prevalence of artists' and content creators' uncritical or contextually negligent use of technologies for various reasons, some of which I outlined within the contexts of computational art in Section 3.4. Furthermore, the aggressive marketing of generative AI as "artistic" marks a decade of the AI industry's strategic targeting of art and creative expression as avenues to culturalize its products, secure its economic interests, and promote its political views. It complements the tech science and industry's implicit sanctioning of relational deficiencies and psychological disorders as acceptable trade-offs of some of their employees' otherwise desirable talents (Dayan 2017; Wayne Meade et al. 2018) and the "justification" of sociopathic tech entrepreneurs by the economic successes of their dashing but morally dubious and socially detrimental business ventures (Jacoby 2020; Marx 2023).

The confluence of computer science and AI industry's questionable presumptions, concepts, and ideological tendencies in generative AI hijacks our cultural intuition (Pedwell 2022), shapes the visions of the future (Scorzin 2023, p. 188), translates into art practices and their public reception (Lossin 2022), channels the professional and popular art discourse, and influences the notions of art and creativity. For instance, the claims of generative AI's mission to "democratize artmaking" chime cyberlibertarian myths about the democratizing powers of markets and digital technologies (Golumbia 2016). The exploitation of evolved human bias toward detecting agency in AI art, its media representation, and public reception reflects the anthropomorphism in AI research. Artists' motives for relegating expressive decisions to generative AI systems converge into hedging or minimizing their responsibilities and foregrounding the benefits of automated cultural production (Browne 2022). The consequent notions of art made by "autonomous" AI entities reinforce the AI industry's sociotechnical blindness. Users' compliance with generative models' censorship and moderation criteria (Riccio et al. 2024) upholds the AI industry's systemic confinement of clients' benefits from leveraging its products.

From the institutional digitization of art collections, through the artists' portfolio websites, to the informal and social media agglomerates of art examples and samples, art stakeholders unwittingly abet the cultural datafication in which their creative work gets misappropriated for training commercial AI models. Their apparently sensible adoption of first-aid tech defenses against this misappropriation, such as data poisoning or style masking, sustains their creative and economic dependence on the continuous arms race between the issues and remedies *within* the AI industry, research, and academia. It inadvertently stokes enthusiasm about AI's potential for transforming art, which plays in tune with the techno-solutionist rhetoric whereby only the tech—but not the regulation of techno-economic power—can save us, and simultaneously diminishes the trust in artmaking as a resilient human faculty.

*5.2. Generative AI, Art, and Democracy*

With the abundance of easy-to-use tools that "bring the AI power to the masses by allowing just about anyone" to become an artist (Sanchez 2023, pp. 19–20; Epstein et al. 2023, p. 4; Parthasarathy 2023), generative AI is broadly presented as a technology which democratizes artmaking. However, this uplifting rhetoric ignores the thorough geopolitical and socioeconomic inequality in accessing AI technologies and positioning oneself as an AI-empowered artist/content creator. By simultaneously obscuring the crucial fact that artmaking is not merely a matter of access to the means of expression and presentation, it merges with the rhetoric of art education, the art market, and cultural industries, which combines the lure of the creative job market with the artist-superstar mythology. It revolves around a photoshopped image of artists as open-minded, unconventional, and uncompromising creative heroes whose authenticity transcends the boredom of conventions and takes the most out of life in a quest for civilizational breakthroughs waiting just beyond the horizon of originality and invention (Stallabrass 2006, pp. 1–7). Mythologizing the art profession is as detrimental for art appraisal as for aspiring artists' careers because it foments false expectations and stimulates idealistic desires for recognition, while nothing characterizes professional artmaking better than uncertainty and precarity.[17]

A related critical view on the promotion of generative AI's art-democratizing role is the creativity cult as one of cognitive capitalism's prominent incentives (Reckwitz 2017; Franklin 2023). Somewhat paradoxically, this promotion makes false claims about limited or worn-out human creativity while simultaneously celebrating AI developers' creative capacities for building creativity-boosting software tools. For instance, in a TED2024 discussion, Demis Hassabis (a CEO and co-founder of Alphabet/Google's DeepMind Technologies) expounds his views about the stalled advancement and implicitly the exhaustion of human intelligence in physics and philosophy, and promotes AI as the ultimate tool for unlocking the secrets of the universe (Hassabis 2024). However, we should acknowledge similar views among computational art pioneers, as evidenced by Abraham Moles' advocacy for investigating the human–machine interaction to "alleviate the poverty of human spirit" (Moles 1962).

If we broadly understand democracy as a system for configuring the governance mechanisms of human associations according to the will and best interests of their constituents, claims about the democratization of artmaking make little sense in principle, regardless of the associated technology. Artmaking is primarily an individualistic or idiomatic human activity and the artworld and art market are inherently competitive and often adversarial institutions, although some art collectives and organizations operate by democratic principles. In the AI industry's discourse, the democratization of artmaking is intimately related to the concepts of democracy and freedom in the free market sense: a tool is more "democratic" if it opens new revenue streams, and the freer the market, the freer the people. However, generative AI is one of the AI industry's product assets that open new revenue streams by destabilizing and potentially killing off many skilled jobs in the arts. Faster development with fewer staff is appealing for illustration, design, gaming, film, and other art-related enterprises where profitmaking takes primacy over the concerns about their human resources' well-being. While automation is useful for relieving unjust, tedious, dehumanizing, and hazardous types of labor (Barley 2020), most forms of creative work supplanted by AI are on the opposite pole of these job categories and some authors argue that the core logic of AI development goes against the respect for humans' intellectual and economic rights (Golumbia 2022; Sætra 2023).

In contrast to the art democratization and creativity enhancement narratives, modern AI's sociopolitical vectors align with and can be traced back to powerful institutions' historical contribution to injustices such as colonizing, uprooting, dispossessing, dehumanizing, and commodifying individuals (Davis et al. 2021; see also Collins and Bilge 2020). The corporate AI and IT cultures are known for dismissing concerns of gender, race, class, and sexual identity and obstructing efforts aimed at breaking down discriminatory barriers. They are equally notorious for ethics-washing: superficially aligning with social justice concerns but discouraging real efforts for deeply reflective and systemic critical approaches and transformative practices (Tacheva and Ramasubramanian 2023).

*5.3. The Elusive Paradigms*

Although the last few years of generative AI development have shown that a human-equivalent AI is still far away, if it is even possible, the AI industry, media, and tech pundits continue juggling "big" concepts, such as intentionality, consciousness, intelligence, sentience, and personhood (Bajohr 2023, p. 59). The arts and humanities academia have acquired a similar taste for hyperbolism, so expressions such as "fundamental changes", "radical disruptions", and "paradigm shifts" frequently adorn the essays on the relationships between AI and art, indicating authors' conflation of art's socioeconomic status with the more basic mental and physical processes in conceiving and making art. Concurrently, post-prefixed neologisms such as "post-AI art" and "art after AI"—associated with the cultural conditions where AI technologies have become ubiquitous—impose a misleading air of discreetness and a false sense of finality onto the gradual coevolution of artistic sensibilities, culture, society, and technology (see Wagenknecht 2018).

As Geoff Cox noted criticizing the term "post-digital art" (Cox 2014), this terminological precession sounds deep when introduced but does not age well and relatively quickly becomes stale facing the contemporaneous realities because the intricacies of art and technologies make the longer-term outcomes and implications of their intersections unpredictable. Summing up her reservations toward the claims of emerging AI technologies' profound breakthroughs in the paper Why AI is Harder than We Think, Melanie Mitchell traced one of the causes for AI's hype/bust cycles to our limited understanding of the complex nature of intelligence itself (Mitchell 2021). Analogously, exaggerations about the "essential" technologically induced transformations of artmaking may be caused by the lack of appreciation or plain disregard for art's subtlety and mutability.

## 6. Conclusions

Like other technologies used in the arts, generative AI imposes the aesthetic, cultural, political, and other norms of its developers and owners onto artists' creative decisions and this lack of exclusive expressive control can be subtle and hard to notice. Therefore, artists' and content creators' uncritical practices can easily become social conduits for the oppressive ideologies of cyberlibertarianism and neo-colonialism. This risk equally emphasizes the need to think about the ways generative AI technologies could be used critically.

In such a context, it seems prudent to be sensitive to the cultural seductiveness of AI and ask how much our critical epistemologies and existential projections (intentionally or not) lean toward the misanthropic sway of AI technoscience. For example, in a wide-reaching treatise titled The Model is the Message (Bratton and Arcas 2022), Benjamin Bratton and Blaise Agüera y Arcas argued that a viable foresight in the AI science and philosophy requires more nuanced analytical, critical, and speculative discussions of central AI issues, terms, and concepts but themselves excluded nuances that do not fit their agendas. To illustrate the narrow scope of current ethical concerns, they cited Timnit Gebru's line: "I don't want to talk about sentient robots, because, at all ends of the spectrum, there are humans harming other humans" but excised its closing: "and that's where I'd like the conversation to be focused" (Johnson 2022), which underlines the emphasis, not the exclusivity, of Gebru's critique. Bratton and Agüera y Arcas also contended that it would be unwise to take the initial, present, or most apparent uses of AI as its ultimate purpose horizon against the fact that AI development and application have been closely and consistently aligned with the core capitalist interests and trends for seventy years (Nadin 2018; Srnicek 2018).[18] The concerns about common-good ethical values are attenuated in their essay and remain peripheral to the ongoing AI conversation in general because they entail facing and addressing the nasty sides of human nature—alienation, parochialism, arrogance, hypocrisy, vanity, greed, subjugation, and discrimination—which fuel the AI industry's evangelistically intoned social politics.

In critical AI studies, the shady undertows of AI-influenced culture need to be addressed more assertively as amalgamations of troublesome economic interests (Dyer-Witheford et al. 2019), self-serving anthropocentrism (Zeilinger 2021), human propensity for deception, self-deception, and cognitive compartmentalization (Trivers 2011), and exploitative virtue signaling (Miller 2019b). The avoidance or euphemistic treatment of unflattering but costly human traits in AI studies is partly caused by the false dichotomy of culture versus biology, whereas culture rests upon and emerges from our evolved mental architecture and cannot be understood without it (Buss 2001). A sincerely introspective reassessment of this architecture should drive the fundamental sociocultural changes we claim to be seeking so much because AI science and industry's shaping of our lives is not the magical force of some cosmic teleology but the cumulative outcome of human motives and actions. It would be reckless to relinquish it to an exclusive social stratum with a proclivity for relational deficiencies, psychological disorders, and abusive ideologies. Instead, it is crucial to curtail the absurd fiction of its inevitability, devise instruments for its meaningful critique, and attain the political will to assume its control.

By understanding the mise-en-scène of AI's cultural sway, we can cultivate an informed and responsible approach to contemporary art and AI. This perspective elucidates the contradictions that infuse our notions about AI in artmaking and adds a touch of skepticism when asking how techno-cultural trends, such as generative AI, transform our relationships with art and in which directions they stir arts' social, economic, and political roles. Further multidisciplinary research is needed to decipher the "symbiotic logic" of the computer science and AI industry's foundational issues and trace the routes of their impact on other cultural domains. It can motivate and inform the inter-institutional cooperation to challenge the dominant cultures of corporate AI and stir toward interventions that empower affected stakeholders (Davis et al. 2021; Tacheva and Ramasubramanian 2023). More importantly, modern sociopolitical principles for handling techno-capital should be systematically tailored and applied beyond specific, imminent, or trendy problems such as AI. The global society must strive not just to regulate the impactful emerging technologies ex post facto, but to actively decide upon and control their functional features and use directions. While appreciating that "common social good" is neither a stable nor universally accepted category, it needs to address the problem of collective political action and mature a more humane economy driven by solidarity, sustainability, quality, reliability, and durability rather than fetishized growth, selfishness, and greed.

**Funding:** This research received no external funding.

**Institutional Review Board Statement:** Not applicable.

**Informed Consent Statement:** Not applicable.

**Data Availability Statement:** Not applicable.

**Conflicts of Interest:** The author declares no conflict of interest.

**Notes**

1.  Epstein et al. (2023), McCormack et al. (2023), Sanchez (2023), and Totlani (2023) provide concise overviews of generative AI and its common issues.
2.  Trained on a dataset of 400 million Internet-scraped image-text pairs, OpenAI's CLIP (Contrastive Language–Image Pre-training) extends the earlier models' object recognition capabilities by retrieving abstract concepts such as context and style (Radford et al. 2021). Other large-scale language-vision models include ALIGN, BASIC, GLIDE, Flamingo, and Imagen.
3.  Diffusion models inject noise interference into data and generate samples in a gradual denoising process that involves predicting the next datum based on prior information found in the dataset. Over time, the model's prediction improves in "filling in the blanks" by calculating the most probable configuration of numerical representations in the data space (see Yang et al. 2022).
4.  Censorship has been integral to the Internet since its outset but has intensified since the mid-1990s with the privatization of the Internet's backbone network, Domain Name System, and the Internet Protocol. It escalated after the release of the Digital Millennium Copyright Act in 1998, and the introduction of Web 2.0 in the mid-2000s (Faris et al. 2008; Cobbe 2021).
5.  Computational creativity is a subfield of computer science that examines AI systems' creative potentials.
6.  For a summary of the common problems in CC studies, see Moruzzi (2020, pp. 162–64; 2022, pp. 183–84) and Issak and Varshney (2022).
7.  The notional realm of "art" in AI research has mostly favored popular artists' paintings from the Western art canon on account of other (visual) artists, art forms, and cultural domains, often interpreted with simplistic categorization taxonomies and anachronistic interpretations of art-historical concepts (Wasielewski 2023). While its scope expanded beyond well-known artists and styles with the large/foundation models, the field's overall approach to art remains superficial.
8.  Most notably the commercialization of personal computers in the 1980s, Apple's doctrine of user-friendly computation in the mid-1980s, the Internet in the 1990s, Web 2.0 technologies in the mid-2000s, blockchain in the late 2000s, and the NFTs and AI media generators since the mid-2010s.
9.  The term art brut ("raw art" or "rough art") was introduced in the 1940s by French artist Jean Dubuffet.
10. Although kitsch is often derogated as culturally detrimental (see Quaranta 2023, pp. 5–6), some authors believe it plays positive social roles by enabling less privileged groups to access art and own the affordable interpretations of artworks, which facilitates a sense of community and promotes wellbeing (Jerrentrup 2024, pp. 4–5).
11. China, Denmark, Finland, France, Germany, Holland, Iceland, Italy, Kenya, Portugal, Russia, Turkey, Ukraine, and the United States.
12. While acknowledging the caveats of retrospective diagnoses, O'Connell and Fitzgerald's (2003) analysis of Turing's biography and contemporaneous accounts concludes that he met Gillberg, ICD-10, and DSM-IV criteria for Asperger's syndrome, which places him within the autism spectrum disorder.
13. For example, see the reiterations of the computer-as-a-person concept by the computer science/AI giant Marvin Minsky in Elis (2014) and vivid accounts of the prevalence of similar mindsets among the Silicon Valley hackers in Jaron Lanier's autobiography Dawn of the New Everything (Lanier 2017).
14. The AI-powered statistical reductionism is not exclusive to businesses and can be radicalized by authoritarian regimes. For example, the Social Credit System and the "innovative development pilot zones", implemented by the Chinese government and AI industry in 2014 and 2019, respectively, are based on a state-wide networked surveillance of citizens' social and business activities with practical repercussions such as the availability of jobs, education, bank loans, electronic services, transportation, and travel (Yang 2022).
15. With large models such as Open AI's GPT-4, this process is somewhat streamlined by using datasets of users' feedback to build separate "training reward" models.
16. Golumbia also monetizes his essays behind the paywall on Medium, while Marx hosts his podcast Tech Won't Save Us on YouTube and monetizes it additionally through Patreon.
17. After enjoying exciting but highly privileged and surreally insulated academic lives, art school graduates quickly learn that talent and hard work do not guarantee success as most of them fail to become emerging artists and only a fraction of those who do maintain mid- or long-term professional careers that can be broadly characterized as artistic.
18. Deeply embedded in and emerging from capitalist profit-seeking motivations that tend to overlook social justice, AI entrepreneurships often gravitate towards free markets, such as the US, to avoid regulation and accountability, and leverage legal loopholes and the absence of ethical vetting (McElroy 2024).